\documentclass{aa}{
\usepackage{txfonts}
\usepackage{graphicx}

\begin{document}
 
\title{Can Microlensing Explain the Long-Term Optical Variability of Quasars?} 
\markboth{E. Zackrisson et al.: }{Can Microlensing Explain the Long-Term Optical Variability of Quasars?} 
\authorrunning{E. Zackrisson et al.}
\titlerunning{Can Microlensing Explain the Long-Term Optical Variability of Quasars?} 

\author{E. Zackrisson\inst{1} \and N. Bergvall\inst{1} \and T. Marquart\inst{1} \and P. Helbig\inst{2}}

\offprints{Erik Zackrisson, 
\email{ez@astro.uu.se}}

\institute{Department of Astronomy and Space Physics, Box 515, S-75120 Uppsala, Sweden \and  Multivax C\&R, Friedrich-Ebert-Str. 83, D-61118 Bad Vilbel, Germany}

\date{Received 10 March 2003 / Accepted 6 June 2003}

\abstract{Although controversial, the scenario of microlensing as the dominant mechanism for the long-term optical variability of quasars does provide a natural explanation for both the statistical symmetry, achromaticity and lack of cosmological time dilation in quasar light curves. Here, we investigate to what extent dark matter populations of compact objects allowed in the currently favored $\Omega_\mathrm{M}=0.3$, $\Omega_\Lambda=0.7$ cosmology really can explain the quantitative statistical features of the observed variability. We find that microlensing reasonably well reproduces the average structure function of quasars, but fails to explain both the high fraction of objects with amplitudes higher than 0.35 magnitudes and the mean amplitudes observed at redshifts below one. Even though microlensing may still contribute to the long-term optical variability at some level, another significant mechanism must also be involved. This severely complicates the task of using light-curve statistics from quasars which are not multiply imaged to isolate properties of any cosmologically significant population of compact objects which may in fact be present.
\keywords{Dark matter -- gravitational lensing -- quasars: general -- cosmology: miscellaneous}}

\maketitle

\section{Introduction} In principle, the possibility that the long-term optical variability of quasars may be due to microlensing caused by a population of planetary-mass compact objects (Hawkins \cite{Hawkins 1993}, Hawkins \cite{Hawkins 1996}) has several attractive features. It provides a natural explanation for the symmetrical and largely achromatic nature of quasar variability on time scales of a few years, in addition to explaining the lack of cosmological time dilation in the variability time scales (Hawkins \cite{Hawkins 2001}). This idea is nonetheless controversial, since it appears to require a cosmologically significant population of dark, compact bodies in a mass range which due to constraints from the MACHO and EROS projects (e.g. Alcock et al. \cite{Alcock et al.}, Lasserre et al. \cite{Lasserre et al.}) can contribute no more than a small mass fraction to the dark halo of the Milky Way. Inferring a limit on the cosmological density of compact objects from the mass fraction of a single halo is, however, difficult given our poor knowledge of the cosmic variance of dark matter properties. There are also concerns that more diffuse lenses (e.g. gas clouds) would not produce sufficiently strong microlensing effects to be detectable in our own halo, but only at cosmological distances (Walker \& Wardle \cite{Walker & Wardle}). Gibson \& Schild (\cite{Gibson & Schild}) have furthermore suggested that strong clustering of lenses inside our halo could make a large population of planetary-mass bodies evade the MACHO/EROS constraints, yet allow detection of such objects in studies of quasar microlensing, where the optical depth due to lensing is much higher.

Several papers critical of the scenario of microlensing-induced quasar variability have been published. Lacey (\cite{Lacey}) \& Alexander (\cite{Alexander}) both modelled the relation between time scale and redshift predicted by microlensing scenarios and found the result to be in conflict with observations of Hawkins (\cite{Hawkins 1993}). The observed relation has since been slightly revised (Hawkins \cite{Hawkins 1996}), and neither Alexander nor Lacey explored the entire parameter space relevant for the microlensing scenario or the selection effects involved. Baganoff \& Malkan (\cite{Baganoff & Malkan}) argued that the observed redshift trend could be explained by intrinsic, wavelength-dependent variability, but this was later refuted by Hawkins \& Taylor (\cite{Hawkins & Taylor}) on the basis of predicting color-dependent variation time scales in conflict with observations.

Despite the qualitative agreement between statistical properties of long-term optical  quasar variability and expectations from microlensing scenarios, the detailed comparisons performed between observations and microlensing simulations have so far been very limited. Here, we use numerical simulations of quasar microlensing to investigate the full parameter space of relevance for the currently favored $\Omega_\mathrm{M}=0.3, \Omega_\Lambda=0.7$ cosmology and investigate to what extent microlensing really can reproduce quantitative features of the observed-long term variability, like the average structure function, the amplitude distribution and the amplitude-redshift relation.

In Sect. 2, the machinery used to simulate the microlensing-induced optical variability of quasars on long time scales is outlined. Section 3 describes the parameter space explored and the method of building synthetic quasar samples used in comparison with observations. In Sect. 4, 5 and 6, the observed average structure function, cumulative amplitude distribution and amplitude-redshift relation are compared to the corresponding results from our microlensing simulations. In Sect. 7 we discuss various assumptions on which this investigation is based. Sect. 8 summarizes our findings.

\section{Numerical simulations of quasar microlensing}
The microlensing model used is described in Zackrisson \& Bergvall (\cite{Zackrisson & Bergvall}), and is based on the multiplicative magnification approximation (Vietri \& Ostriker \cite{Vietri & Ostriker}). In this case, the magnification, $\mu_\mathrm{tot}$, due to $i$ microlenses is equal to the product of the individual ones:
\begin{equation} 
\mu_\mathrm{tot}=\prod\limits_{i}\mu_i.
\label{eq1}
\end{equation}
Although this approximation cannot be expected to reproduce detailed features of microlensing light curves, it has been proved useful (e.g. Pei \cite{Pei 1993}) for statistical investigations and significantly reduces the calculation time required by ray-tracing and similar methods. It has also previously been applied (Schneider \cite{Schneider}, Lacey \cite{Lacey}) to investigations of quasar light-curve statistics. 

When calculating the time-dependent magnification $\mu(t)$ resulting from each compact object, we assume a circular source with uniform brightness and the approximate magnification formulas for point-mass lenses of Schneider (\cite{Schneider}) and Surpi et al. (\cite{Surpi et al.}), relevant for the small and large source regimes respectively. The compact objects acting as microlenses are assumed to be uniformly and randomly distributed throughout the simulated observing beam and the influence of external shear is neglected.  Since the quasars used in this investigation are all located in the direction $\alpha =21^\mathrm{h} 28 ^\mathrm{m}$, $\delta=-45\degr$, a universal observer velocity will be used. Adopting the velocity of the Sun relative to the cosmic microwave background derived by Lineweaver et al. (\cite{Lineweaver et al.}), the velocity of the observer perpendicular to the line of sight becomes $307.9$ km/s. The velocities of quasars perpendicular to the line of sight are assumed to be normally distributed with velocity dispersion $\sigma_\mathrm{v,QSO}=300$ km/s. 

Unless otherwise stated, all calculations will be carried out in the framework of the currently favored $\Lambda$-dominated cosmology, with $\Omega_\mathrm{M}=0.3$, $\Omega_\Lambda=0.7$ and H$_0=65$ km/s/Mpc. The mean density inside the simulated observing beam is assumed to be equal to the mean density of the universe, which implies a homogeneity parameter $\eta=1.0$ (Kayser et al. \cite{Kayser et al.}). 

Zackrisson \& Bergvall (\cite{Zackrisson & Bergvall}) may be consulted for further details on the method used.

\section{Synthetic quasar samples} 
In order to be able to draw firm conclusions about possible discrepancies between simulations and observations, it is important to probe the entire plausible range of microlensing parameter values. When comparing statistics derived from samples of observed and simulated light curves, potential selection effects must also be considered.  

\subsection{The microlensing parameter space}
For each combination of microlensing parameter values listed in Table \ref{parameters}, 35000 light curves have been generated, uniformly distributed in the range of quasar redshifts $z_\mathrm{QSO}=0.13-3.6$, giving on the order of 15 million light curves. Here, $\Omega_\mathrm{compact}$ represents the contribution to the cosmological density from the compact object population, $M_\mathrm{compact}$ the mass of the compact objects responsible for the microlensing effect, $\sigma_\mathrm{v,compact}$ the velocity of compact objects perpendicular to the line of sight and $R_\mathrm{QSO}$ the typical size of the optical-UV continuum-emitting region of quasars, normally assumed to correspond to some fraction of an accretion disc surrounding a central supermassive black hole. In this investigation, we will assume all compact objects to have the same $M_\mathrm{compact}$ and all quasars to have the same $R_\mathrm{QSO}$. Even though more complicated mass spectra and distributions of source sizes may be present in reality, we will show that this has no impact on the conclusions of this paper.

We believe that these simulations span the entire microlensing parameter space relevant for the present situation. The cosmological density of compact objects, $\Omega_\mathrm{compact}$, is limited upwards by the adopted background cosmology ($\Omega_\mathrm{M}=0.3$, $\Omega_\Lambda=0.7$), whereas at $\Omega_\mathrm{compact} < 0.05$ the variations induced become much too small to explain the observed quasar variability. The time scale of variations induced by lens masses higher than $M_\mathrm{compact}=1 \ M_\odot$ would typically be longer than the time span of present monitoring campaigns, and lens masses lower than $10^{-5} \ M_\odot$ would give negligible variations except in the case of the very smallest $R_\mathrm{QSO}$. Where in doubt, single samples at $10^{-6} \ M_\odot$ have been generated for test purposes. The velocity dispersion of the compact objects is related to their preferred whereabouts, i.e. whether mainly located in the field or in rich clusters of galaxies. In Zackrisson \& Bergvall (\cite{Zackrisson & Bergvall}) we argue, based on results from cold dark matter simulations, that $\sigma_\mathrm{v,compact}=200$--600 km/s represents the likely velocity span. The parameter $R_\mathrm{QSO}$ is not well determined observationally, but in the framework of accretion-disc theory constrained to be at least a few times larger than the gravitational radius of the central supermassive black hole, which for a $10^{8-9} \ M_\odot$ objects amounts to $1.5\times 10^{11-12}$ m. The smallest value considered here is $R_\mathrm{QSO}=10^{12}$ m. Values of $R_\mathrm{QSO}$ significantly larger than $3\times 10^{13}$ m give very modest light variations and are therefore not of interest for the present investigation.
\begin{table}[h]
\caption[]{The set of discrete microlensing parameter values which define the grid of synthetic light-curve samples used in this study.} 
\begin{flushleft}
\begin{tabular}{lllllll} 
\hline
\hline
$\Omega_\mathrm{compact}$: & 0.05 & 0.1 & 0.15 & 0.2 & 0.25 & 0.3\cr
$\log M_\mathrm{compact} \ (M_\odot)$: &-5 & -4 & -3 & -2 & -1 & 0 \cr 
$\sigma_\mathrm{v,compact} \ (\mathrm{km/s})$: & 200 & 400 & 600 & & &\cr
$\log R_\mathrm{QSO} \ (\mathrm{m})$: & 12 & 12.5 & 13 & 13.5 & &\cr
\hline
\end{tabular}
\end{flushleft}
\label{parameters}
\end{table}

\subsection{Light curves}
Each microlensing light curve generated has a time span of 25 years, where the time-dependent magnification is estimated four times per year within a period of three months. The intrayear variation in magnitudes is then averaged over to give 25 magnification data points $m(t)$, where $m(t)=-2.5\log\mu(t)$, in close agreement with the procedure used in the observed quasar light curves used for comparison. 

Averaging over the intrayear magnification data points has the effect of smearing variations that occur on time scales smaller than three months and reducing the amplitudes for microlensing scenarios where a significant fraction of the variations are of this kind. Although the intrayear observations were sometimes carried out within a period shorter than assumed here, and in some years only one plate was available, tests using no magnitude averaging in the yearly data points of the microlensing light curves do however show that this discrepancy between model and observations has no significant impact on the conclusions of this paper.

In Hawkins (\cite{Hawkins 1996}), the variability on time scales shorter than one year was studied by taking 16 plates over a period of six months. Small short-term variations, supposedly intrinsic in nature, were noted. Even though the present study is focused on the long-term variability of quasars, such short-term variations will give a small contribution to the variations between the yearly averages, and therefore need to be considered when assessing the reason for possible discrepancies between the observed long-term variability and that predicted by microlensing. The observed dispersion among the roughly four plates taken each year amounts to 0.13 magnitudes (Hawkins 2003, private communication), where part of the variability comes from the approximately 0.08 magnitude measurement error. Assuming the variations due to short-term variability to be Gaussian and uncorrelated, the effect of both measurement error and intrinsic short-term variability can be modelled by adding a $\sigma_m=0.065$ component of Gaussian noise to each yearly $m(t)$. This procedure, which we adopt in the following, is however not entirely consistent for the whole microlensing parameter space probed, since many combinations of small $M_\mathrm{compact}$ and small $R_\mathrm{QSO}$ already predict variability among the four intrayear measurements on a level close to that attributed to the intrinsic component, and may therefore slightly overpredict the variability of such scenarios. This will however not have any impact on the outcome of the present investigation. 

\subsection{Light-curve statistics}
The comparison between observed and simulated light curves will be carried out in terms of the average structure function and the light curve amplitude. The average first-order structure function of a sample of microlensing light curves, representing a curve of growth of variability with time lag, is defined as:
\begin{equation}
S(\tau)=\sqrt{\left<\frac{1}{N(\tau)}\sum_{i<j}\left[m(t_j)-m(t_i)\right]^2\right>},
\label{eq2}
\end{equation}
where the sum runs over the $N(\tau)$ data points for which $t_j-t_i=\tau$. 

Observationally, the amplitude $\delta m$ is defined as the difference between the minimum and maximum magnitude observed in a quasar when monitored for a certain period of time with a particular sampling rate. For the case of variations induced purely by microlensing, this reduces to:
\begin{equation} 
\delta m=\max (m(t))-\min (m(t)). 
\label{eq3}
\end{equation}

In the present investigation, the amplitude is derived from the first 21 data points of the light curve, in accordance with the observational procedure of Hawkins (\cite{Hawkins 2000}). 

\subsection{Selection effects}
When comparing the results from microlensing simulations to observations, we will use light-curve statistics derived from the UVX sample of Hawkins (\cite{Hawkins 2000}), containing 184 objects brighter than  $m_\mathrm{B}=21.5$ in the redshift range $z_\mathrm{QSO}=0.242-2.21$, which has been selected on basis of ultraviolet excess ($U-B < -0.2$). Despite its smaller redshift span, the relevant selection effects are more easily simulated than in the case of variability-selected samples, where the proposed variability mechanism must generate approximately the same fraction of amplitudes higher than the sample threshold as in real quasars in order not to introduce unrealistic biases. As will be shown in Sect. 5, microlensing by itself does not seem to be able to fulfill this criterion. 

Assuming the colour-criterion of the UVX sample not to correlate with variability (see Sect. 7 for a discussion), the most important selection effect to consider is the amplification bias. A quasar that lacks sufficient intrinsic brightness to be included in a flux-limited sample may still be magnified through gravitational microlensing to reach above the threshold for detection. A flux-limited sample will therefore contain an enhanced fraction of highly magnified objects. As demonstrated in Schneider (\cite{Schneider}), magnification and amplitude correlate, implying that a flux-limited sample should also display an enhanced probability for large amplitudes. This effect may be further boosted by the fact that each data point in the observed light curve is also associated with an observational error, which statistically increases the light curve amplitudes. Provided that the error in the data point used to define the magnitude of an object is correlated with the amplitude of its light curve, an effect similar to that of amplification bias will result from the measurement noise alone, increasing the probabilities of high amplitudes after the magnitude limit has been imposed. In the present case, however, we find that the measurement error is too small and the number of data points in the light curve too large for this measurement uncertainty bias to have any significant effect on the statistical variability properties of the sample.

The effects of amplification bias is simulated by assigning each microlensing light curve a random absolute magnitude following the probability predicted by the optical luminosity function of Boyle et al. (\cite{Boyle et al.}), converting this into an apparent magnitude given the background cosmology and assuming a power-law continuum ($L_\nu \propto \nu^{-\alpha}$) with slope $\alpha=0.5$ for all quasars when calculating the $k$-correction. The magnification, $m(0)$, combined with the corresponding $\sigma_m=0.065$ short-term variability (measurement error and intrinsic intrayear variability) of the first data point in the light curve is then added to the magnitude. Subsequently, all light curves falling on the faint side of the UVX magnitude limit are rejected. A selection is finally made from the remaining light curves to ensure that the redshift and magnitude distributions of the simulated quasar sample fully match that of the observed one.

\section{The average structure function} 
\begin{figure}[t]
\resizebox{\hsize}{!}{\includegraphics{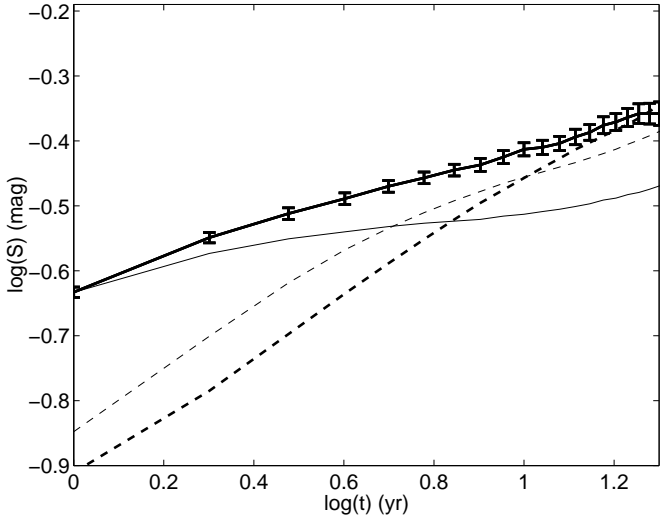}}
\caption[]{The observed UVX average structure function (thick solid) and the best-fitting (in a least-square sense) average structure functions predicted by microlensing scenarios with parameter constraints $\Omega_\mathrm{compact}\leq 0.15$ (thin solid),  $M_\mathrm{compact}\geq 10^{-1} \ M_\odot$ (thick dashed), and $R_\mathrm{QSO}\geq 3\times10^{13}$ m (thin dashed). All simulated average structure functions obeying these parameter constraints but having larger residuals are located below the ones plotted.}
\label{SF_paramtest}
\end{figure}
The observed first order structure function of quasars typically consists of three distinct parts (Hughes et al. \cite{Hughes et al.}). For time lags longer than the longest correlation time scale, there is a plateau at a value equal to twice the variance of the fluctuation. For short time lags, there is another plateau at twice the variance of the measurement noise. These regions are linked by a curve whose slope is related to the nature of the variability of the source. In Hawkins (\cite{Hawkins 2002}) the observed power-law slope ($0.20\pm 0.01$) of the average structure function was compared to predictions from three proposed mechanisms of variability: The starburst model, accretion-disc instabilities and microlensing. The close resemblance to the slope predicted by microlensing ($0.25\pm 0.03$), compared to the much steeper slopes predicting by the disc instability ($0.44\pm0.03$) and starburst ($0.83\pm0.08$) scenarios appeared to favor microlensing as the dominant mechanism. The parameter space of the different models was however not probed at any depth. In the case of microlensing, structure functions derived from single microlensing light curves in the literature were used, not average structure functions derived from simulated samples. The effects of the sample selection bias, the distribution of quasar redshifts and the statistical variations between different microlensing light curves were thereby neglected.

Due to the complicated selection procedure of the quasar sample from which the observed average structure function of Hawkins (\cite{Hawkins 2002}) was derived, we will instead compare average structure functions derived from our simulated samples to an observed average structure function (Hawkins 2003, private communication) derived from the more easily reproduced UVX sample. The power-law slope of this average structure function is 0.19, very close to the value derived in Hawkins (\cite{Hawkins 2002}).
\begin{figure}[t]
\resizebox{\hsize}{!}{\includegraphics{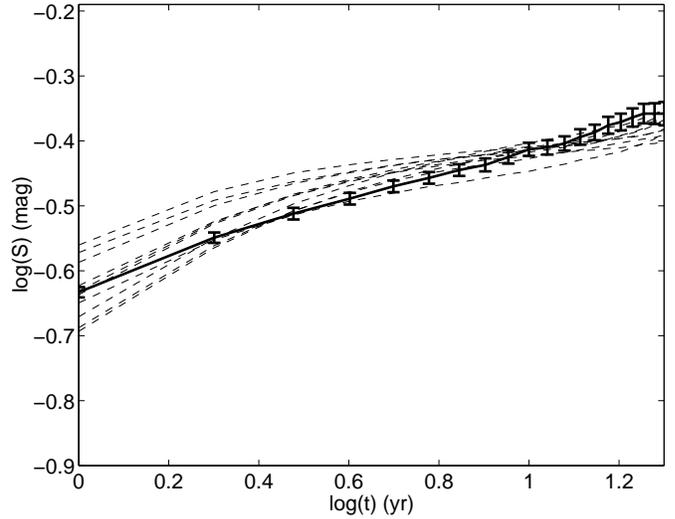}}
\caption[]{The observed UVX average structure function (thick solid) and the ten best-fitting average structure functions derived from microlensing scenarios (dashed).}
\label{SF_bestfits}
\end{figure}

Apart from the slope, the scaling of the structure function is also important, since this is related to the variance of the light curves. It turns out that most of the simulated average structure functions lie significantly below the observed one, indicating that most parameter combinations generate too little variability. If microlensing is to be able to reproduce the observed average structure function, certain regions of the microlensing parameter space can therefore be ruled out. In Fig.~\ref{SF_paramtest}, we plot the observed average structure function in comparison to the average structure functions with the highest scaling out of those predicted by microlensing in specific parameter ranges. On this basis, all scenarios with $\Omega_\mathrm{compact}\leq 0.15$, $M_\mathrm{compact}\geq 10^{-1} \ M_\odot$ or $R_\mathrm{QSO}\geq 3\times 10^{13}$, out of those allowed by the microlensing parameter grid defined in Table 1, can be ruled out as inadequate. Scenarios with $M_\mathrm{compact}=10^{-5} \ M_\odot$ can provide a high degree of variability but have a maximum average structure function slope of 0.06 (indicating too much of variability occurring on short time scales) and are therefore also unsuitable candidates.

In Fig.~\ref{SF_bestfits} we plot the observed average structure function derived from the UVX sample and the ten best-fitting (in a least-square sense) simulated average structure functions. Even though no single microlensing curve completely reproduces the shape and scaling of the observed average structure function, the agreement is reasonable given the limited resolution of the explored microlensing parameter space. The parameters of the best fits lie inside the range $\Omega_\mathrm{compact}=0.2$--0.3, $M_\mathrm{compact}=10^{-4}$--$10^{-3}\ M_\odot$, $\sigma_\mathrm{v, compact}=200$--600 km/s and $R_\mathrm{QSO}=10^{12}$--$10^{13}$ m. This method of determining the lens masses required by any viable scenario of microlensing-induced long-term quasar variability is in good agreement with that of Hawkins (\cite{Hawkins 1996}), where the mass of the lensing objects was estimated at $\approx 10^{-3} \ M_\odot$ using simple time-scale arguments.

When fitting power-law slopes to the 432 average structure functions derived from simulated samples based on the different microlensing parameter configurations of Table \ref{parameters}, we find the slope to vary in the range 0.004--0.50. The slope of 0.25 derived in Hawkins (\cite{Hawkins 2002}) from single microlensing light curves in the literature can therefore not be said to be a robust prediction of the microlensing mechanism. Instead, the fitted slope is quite sensitive to the model parameters assumed. Even in the parameter region spanned by the best average structure function fits, the range of slopes is 0.06--0.24. 

Even though a large part of the $\Omega_\mathrm{compact}=0.2$--0.3, $M_\mathrm{compact}=10^{-4}$--$10^{-3}\ M_\odot$, $R_\mathrm{QSO}=10^{12}$--$10^{13}$ m microlensing parameter space can actually be ruled out on the basis that such scenarios significantly overpredict the fraction of high amplitudes (Zackrisson \& Bergvall \cite{Zackrisson & Bergvall}), we make no attempt here to determine how close the observed average structure function can be reproduced in the absence of these, since this would require a significant increase in the parameter resolution. Instead, we are content to conclude that there is no strong reason to doubt that the observed average structure function can be explained by certain combinations of microlensing parameters. As we will show, this is however not true for the cumulative amplitude distribution or the amplitude-redshift relation.

\section{The cumulative amplitude distribution}
\begin{figure}[t]
\resizebox{\hsize}{!}{\includegraphics{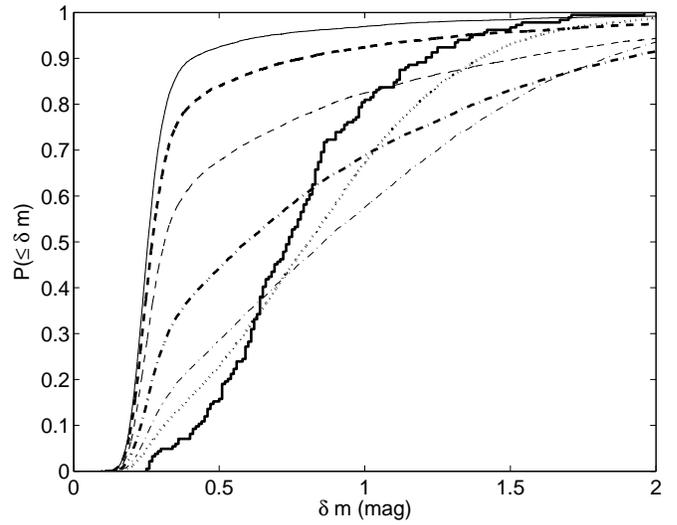}}
\caption[]{The cumulative probabilities $P(\leq \delta m)$ of observing quasars with amplitudes lower than $\delta m$ for the UVX sample (thick solid). The other lines represent the corresponding distributions derived from microlensing simulations assuming $\Omega_\mathrm{compact}=0.3$, $R_\mathrm{QSO}=10^{12}$ m, $\sigma_\mathrm{v,compact}=400$ km/s and $M_\mathrm{compact}=1 \ M_\odot$ (thin solid), $10^{-1} \ M_\odot$ (thick dashed), $10^{-2} \ M_\odot$ (thin dashed), $10^{-3} \ M_\odot$ (thick dash-dotted), $10^{-4} \ M_\odot$ (thin dash-dotted), $10^{-5} \ M_\odot$ (thick dotted).}
\label{CDF1}
\end{figure}
\begin{figure}[t]
\resizebox{\hsize}{!}{\includegraphics{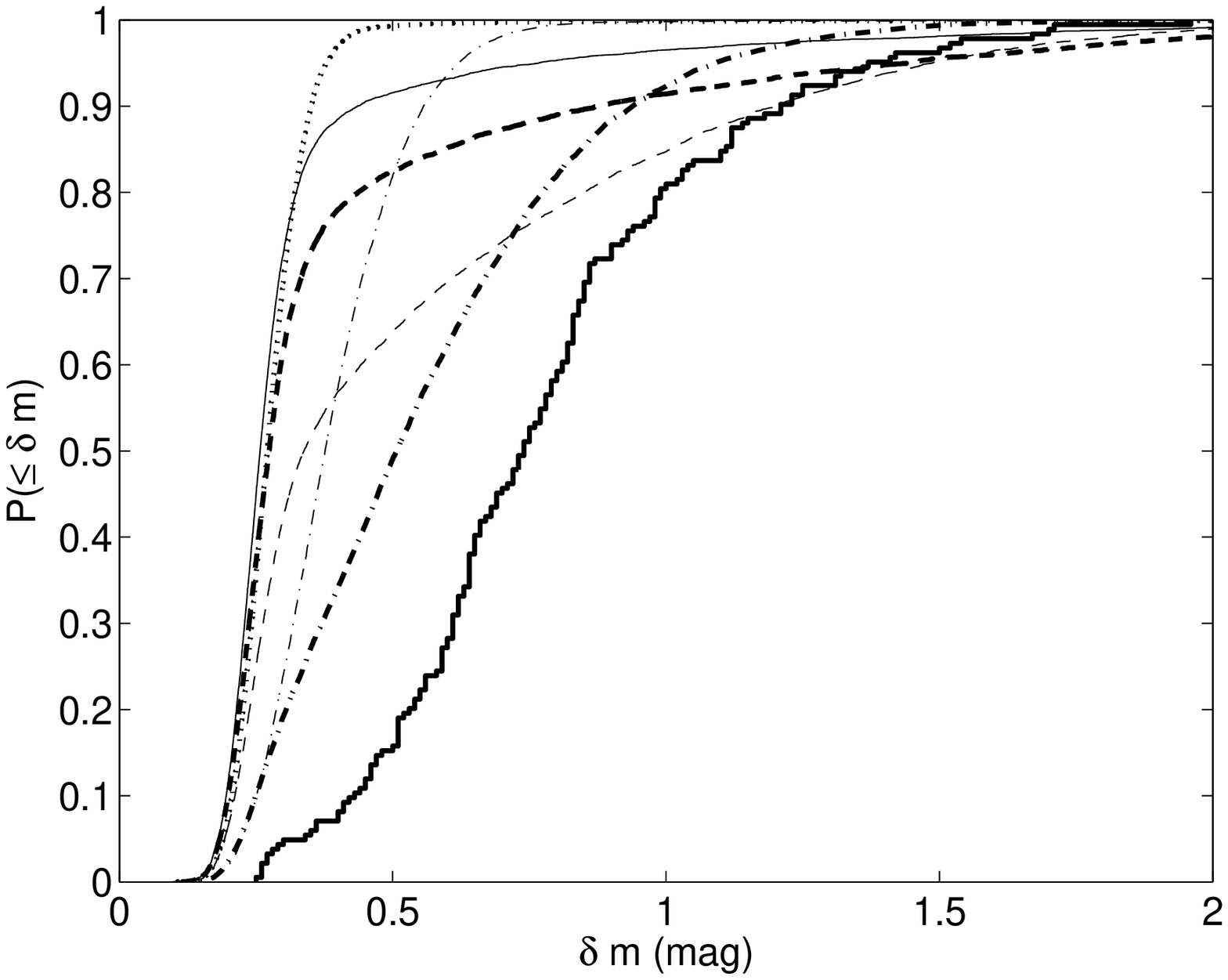}}
\caption[]{Same as Fig.~\ref{CDF1} except $R_\mathrm{QSO}=3\times 10^{13}$ m.}
\label{CDF2}
\end{figure}

In the absence of any variability-related bias in addition to those treated in Sect. 3.4, we should expect microlensing scenarios to approximately reproduce the observed distribution of amplitudes in the UVX sample, provided that microlensing truly is the dominant mechanism for the long-term optical variability of quasars. As shown in Schneider (\cite{Schneider}), the cumulative amplitude distributions predicted by microlensing simulations are very sensitive to the model parameters. In Figs.~\ref{CDF1} and ~\ref{CDF2} we show examples of the cumulative amplitude distributions predicted for different lens masses in the case of $R_\mathrm{QSO}=10^{12} \ \mathrm{m}$ and $3\times 10^{13}\ \mathrm{m}$, respectively.

In Hawkins (\cite{Hawkins 1996}), it was argued that the amplitude distributions predicted by the microlensing simulations of Schneider (\cite{Schneider}), which assumed $R_\mathrm{QSO}=10^{13}$ m and an Einstein-de Sitter Universe, could probably provide a good match with the observed distribution of amplitudes for $M_\mathrm{compact}=10^{-3} \ M_\odot$, provided that $R_\mathrm{QSO}=3\times 10^{13}$ m instead. Our model confirms this to be the case for the Einstein-de Sitter cosmology, but not, as we show in Fig.~\ref{CDF2}, for the $\Omega_\mathrm{M}=0.3$, $\Omega_\Lambda=0.7$ cosmology assumed here. At such large $R_\mathrm{QSO}$, no microlensing scenario even comes close to reproducing the observed amplitude distribution. This is in good agreement with our conclusions from the analysis of average structure functions, where scenarios assuming $R_\mathrm{QSO}\ge 3\times 10^{13} $m were also ruled out on the basis of giving to little variability.

In Fig.~\ref{CDF1}, many of the cumulative probability distributions predicted by microlensing scenarios can be seen to clearly cross over to the right-hand side of the corresponding observed distribution, indicating an overprediction of the fraction of the very highest amplitude objects. This fact was used in Schneider (\cite{Schneider}) and Zackrisson \& Bergvall (\cite{Zackrisson & Bergvall}) to infer upper limits on $\Omega_\mathrm{compact}$ for different $M_\mathrm{compact}$. Here, we will instead concern ourselves with another discrepancy between the observed and predicted amplitude distributions more directly related to the question of whether microlensing really is responsible for the observed light variations or not. 

The selection criterion of $\delta m > 0.35$ has proved useful (Hawkins \cite{Hawkins 2000}) for building observational samples of quasars extending to redshifts higher than the $z_\mathrm{QSO}\approx 2.2$ limit of samples based only on ultraviolet excess. The fraction of objects with amplitudes higher than 0.35 magnitudes predicted by the dominating mechanism of variability therefore has important implications, since it is related to the completeness of such variability-selected samples. If an adequate fraction of amplitudes higher than 0.35 magnitudes cannot easily be reproduced by microlensing, there is no way to simulate the properties of variability-selected samples using the $\delta m > 0.35$ criterion (like in the VAR sample of Hawkins \cite{Hawkins 2000}), since this would mean rejecting an unrealistically large fraction of simulated light curves.

As can be seen in Figs.~\ref{CDF1} and ~\ref{CDF2}, none of the microlensing scenarios explored manage to reproduce the observed fraction of amplitudes higher than 0.35 magnitudes. The microlensing simulations instead predict large numbers of objects below this limit. This turns out to be a general feature of our simulations, and not just of the small number of parameter combinations used in these plots. It should be noted that the absence of amplitudes smaller than 0.2 magnitudes in the predicted distributions is almost entirely due to the presence of simulated observational errors and intrinsic intrayear variability in each data point of the light curves. Without the added short-term noise, significant fractions of even lower amplitudes would be predicted.

\begin{figure}[t]
\resizebox{\hsize}{!}{\includegraphics{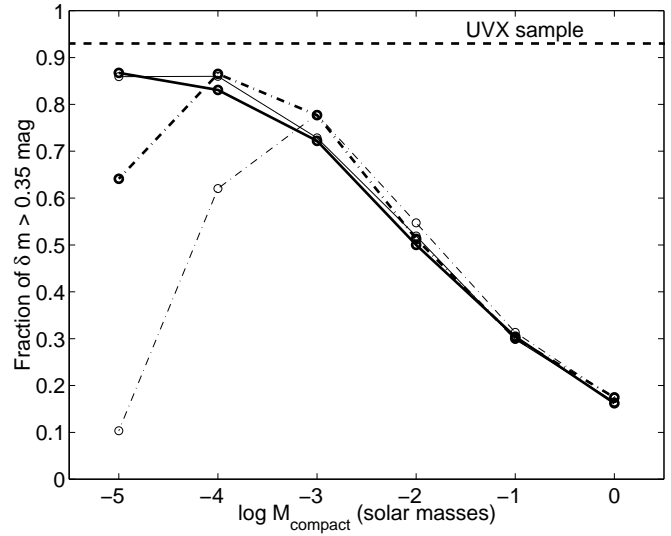}}
\caption[]{The fraction of objects with amplitudes higher than 0.35 magnitudes predicted by microlensing scenarios for different $M_\mathrm{compact}$, assuming $\Omega_\mathrm{compact}=0.3$, $\sigma_\mathrm{v,compact}=600$ km/s and $R_\mathrm{QSO}=10^{12}$ m (thick solid), $3\times 10^{12}$ m (thin solid), $10^{13}$ m (thick dash-dotted), $3\times 10^{13}$ m (thin dash-dotted). The thick dashed line marks the observed fraction (93\%) of the UVX quasar sample.}
\label{fracpic}
\end{figure}

In Fig.~\ref{fracpic} we compare the observed  fraction of objects with amplitudes higher than 0.35 magnitudes in the UVX sample with the corresponding microlensing predictions as a function of $M_\mathrm{compact}$ and $R_\mathrm{QSO}$ in the case of $\Omega_\mathrm{compact}=0.3$ and $\sigma_\mathrm{v,compact}=600$ km/s (the scenario giving the most variability). In no case can microlensing fully reproduce the observed fraction. Given the trend of variability with lens mass in the case of $R_\mathrm{QSO}=10^{12}$ m, one could however imagine that extending the investigation to lower lens masses than $10^{-5} \ M_\odot$ might improve the agreement, but tests show that the predicted fraction of amplitudes above 0.35 magnitudes does not become significantly higher at $10^{-6} \ M_\odot$. Such scenarios can also be ruled out on the basis of giving too much short-term variability (structure function slopes much flatter than observed). 

Hawkins (\cite{Hawkins 2002}) shows that the Seyfert galaxies in his sample are dominated by a different kind of variability. Since the UVX sample used here also contains a number of such objects, one might expect that this could affect the comparison. However, following the procedure of Hawkins (\cite{Hawkins 2002}) of removing objects with $z_\mathrm{QSO}<0.5$ and $M_\mathrm{B}>-23$ to get a clean sample of quasars does not improve the agreement, but actually makes it slightly worse.

We conclude that the fraction of amplitudes higher than 0.35 magnitudes cannot fully be reproduced by our present microlensing simulations. The gap between observations and predictions is however quite small for lenses in the range $M_\mathrm{compact}=10^{-5}$--$10^{-4} \ M_\odot$, as seen in Fig.~\ref{fracpic}.

\section{The amplitude-redshift relation} 
In Hawkins (\cite{Hawkins 2000}), the observational relation between mean amplitude and redshift was explored. In the UVX sample, the mean amplitude was found to increase up to a redshift around one, followed by a roughly constant mean amplitude towards higher redshift. This was attributed to a magnitude effect in the sense of higher luminosity quasar having smaller amplitudes.

The most naive expectation from microlensing scenarios would perhaps be a mean amplitude which continuously increases with redshift due to increased optical depth to lensing. Qualitatively, the observed behaviour of the mean amplitude may however be explained by microlensing-induced variations, provided that the size of the continuum-emitting region at the rest wavelength observed is larger for the more luminous quasars found at high redshifts. 

\begin{figure}[t]
\resizebox{\hsize}{!}{\includegraphics{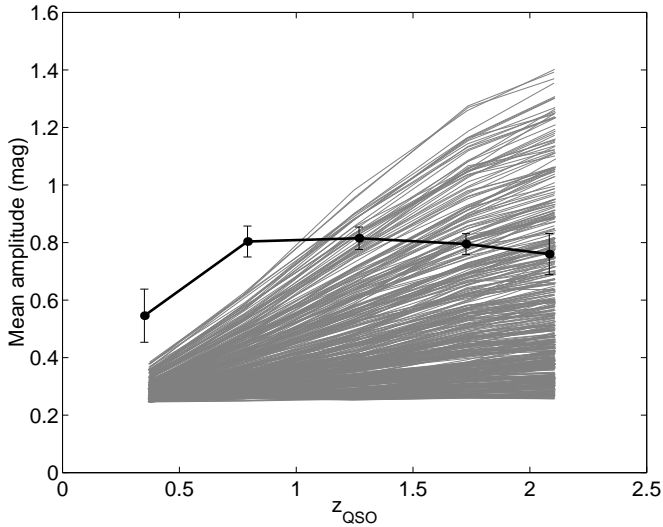}}
\caption[]{The empirical relation between mean amplitude and redshift for the UVX quasar sample (thick black) and the corresponding relations derived from the 432 different microlensing parameter combinations generated (thin grey). The error bars indicate the uncertainty in the position of the observed mean.}
\label{ampzrel}
\end{figure}

In Fig.~\ref{ampzrel}, the empirical amplitude-redshift relation is plotted together with the corresponding relations derived from the 432 different microlensing parameter configurations allowed by Table \ref{parameters}. Since every microlensing curve assumes $R_\mathrm{QSO}$ to be constant with redshift, the fact that the shapes of the simulated and observed relations are very different does not rule out microlensing as a plausible mechanism. More serious is however the fact that no microlensing scenario appears capable of explaining the high mean amplitudes observed at $z_\mathrm{QSO}<1$, where one third of the quasars in the UVX sample are located. Even though relaxing the approximation of constant $R_\mathrm{QSO}$ may, in principle, create a plateau in the amplitude-redshift relation, the mean amplitudes can never lie higher than the maximum microlensing prediction of Fig.~\ref{ampzrel}, provided that microlensing really is the dominating mechanism for variability. This discrepancy  between observations and simulations is much larger than that discovered in connection to the cumulative amplitude distributions. We therefore conclude that, at least for quasars at $z_\mathrm{QSO}<1$, significant contributions to the long-term variability must come from a mechanism other than microlensing in order to explain the observed amplitude-redshift relation. 

\section{Discussion}
Even though our microlensing simulations appear capable of reproducing the observed average structure function of quasars, they fail to explain both the observed fraction of objects with amplitudes higher than 0.35 magnitudes and the high mean amplitudes observed at $z_\mathrm{QSO}<1$. This does not rule out the possibility that microlensing may still contribute to the variability at some level, but indicates that another mechanism must also contribute significantly. 

Since any mechanism other than microlensing contributing to the observed variability would have to give largely achromatic and symmetrical variations to be consistent with observations, this means that microlensing cannot be the only mechanism involved possessing these properties. The exciting possibility of deriving properties of dark matter by comparing results from microlensing simulations to the observed properties of quasar light curves is thereby severely obstructed, since any microlensing signal that may be present in the light-curve statistics is likely to be blended with a different kind of variability with similar characteristics. Although the chromaticity of proposed mechanisms of intrinsic variability  like accretion-disc instabilities and supernova explosions have not been thoroughly investigated, both of these mechanisms appear capable of generating statistically symmetric variations on long time scales (e.g. Kawaguchi et al. \cite{Kawaguchi et al.}, Aretxaga \cite{Aretxaga}). The one piece of evidence pointing uniquely in the direction of an extrinsic mechanism is however the lack of cosmological time dilation in the characteristic time scales of the variations (Hawkins \cite{Hawkins 2001}). Provided that no efficient extrinsic variability mechanism other than microlensing can be found, one possible way of setting a lower limit to the contribution from microlensing could therefore be to investigate how large the microlensing population would have to be to eradicate the cosmological time dilation signal believed to be a feature of the intrinsic variability component.
  
How robust are then the results presented here? So far we have considered only populations of compact objects and quasars characterized by single values of $R_\mathrm{QSO}$ and $M_\mathrm{compact}$. In reality, extended distributions of quasar radii (possibly correlated with quasar luminosities) and lens masses may of course be present. In the framework of the microlensing model used, such distributions pose no threat to our conclusions, as long as our simulations truly probe the entire range of plausible parameter values. Since no single parameter combination explored can reproduce the high mean amplitudes observed at low redshift, neither will any mixture.

\begin{figure}[t]
\resizebox{\hsize}{!}{\includegraphics{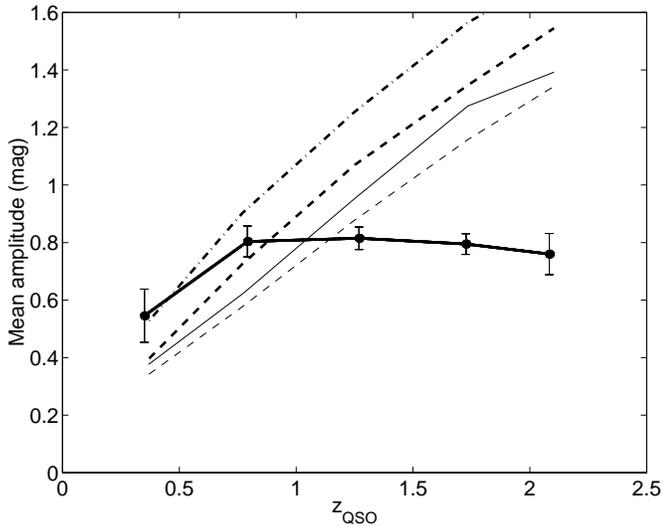}}
\caption[]{The empirical relation between mean amplitude and redshift for the UVX quasar sample (thick solid) and the corresponding predictions from microlensing simulations assuming different background cosmologies: The standard $\Omega_\mathrm{M}=0.3$, $\Omega_\Lambda=0.7$ case (thin solid),   $\Omega_\mathrm{M}=0.4$, $\Omega_\Lambda=0.6$ (thick dashed), $\Omega_\mathrm{M}=0.2$, $\Omega_\Lambda=0.8$ (thin dashed) and $\Omega_\mathrm{M}=1.0$, $\Omega_\Lambda=0.0$ (thick dash-dotted). All simulations assume $\Omega_\mathrm{compact}=\Omega_\mathrm{M}$, $M_\mathrm{compact}=10^{-4} \ M_\odot$, $R_\mathrm{QSO}=10^{12}$ m and $\sigma_\mathrm{v,compact}=400$ km/s. The error bars indicate the uncertainty in the position of the observed mean.}
\label{ampzrel_cosmotest}
\end{figure}

The results presented assume $\Omega_\mathrm{M}=0.3$, $\Omega_\Lambda=0.7$, but the conclusions are not sensitive to minor variations in the adopted cosmology. In Fig.~\ref{ampzrel_cosmotest}, we present the predicted mean amplitude-redshift relation in the case of $\Omega_\mathrm{compact}=\Omega_\mathrm{M}$, $M_\mathrm{compact}=10^{-4} \ M_\odot$, $R_\mathrm{QSO}=10^{12}$ m, $\sigma_\mathrm{v,compact}=400$ km/s (one of the scenarios giving the highest mean amplitude in the lowest redshift bin of Fig.~\ref{ampzrel}) for a number of different combinations of $\Omega_\mathrm{M}$ and $\Omega_\Lambda$. Clearly, variations like ($\Omega_\mathrm{M}=0.2$, $\Omega_\Lambda=0.8$) and ($\Omega_\mathrm{M}=0.4$, $\Omega_\Lambda=0.6$) allowed by current uncertainties in these parameters do not provide an obvious way around the discrepancies between simulations and observations. Only in a cosmology with a matter density significantly higher, like in the case of an Einstein-de Sitter Universe ($\Omega_\mathrm{M}=1.0$, $\Omega_\Lambda=0.0$), is there a reasonable agreement at the lowest redshifts. 

Our simulations also assume $\eta=1$, which is admittedly not self-consistent for the entire microlensing parameter space investigated. This parameter describes the fraction of homogeneously distributed matter, where the scale of clumps in the remaining matter per definition is on the same order of magnitude as the angles involved (Kayser et al. \cite{Kayser et al.}). The $\eta$ parameter affects the angular diameter distances going into the magnification contribution from each lens and the luminosity distances used to make the conversion from absolute to apparent magnitudes. When the microlensing source is large and the variations are due to Poisson fluctuations in the number of compact objects in front of the source (Surpi et al. \cite{Surpi et al.}), rather than due to rare and isolated microlensing events, the mean density of the simulated observing beam becomes approximately equal to the mean density of the universe, which makes $\eta=1$ reasonable. However, when the angular extent of the source is small compared to that of the lenses, the density in the simulated observing beam varies violently with time. In this case $\eta <1$ is more appropriate. This situation arises naturally when modelling the gravitational lensing of supernovae (e.g. M\"ortsell \cite{Mörtsell}), which are generally assumed to be less extended than quasars, but also in the present investigation when small values of $R_\mathrm{QSO}$ are considered. By testing different values of $\eta$ in the range $\eta=0$--$1$, we have however found that variations in this parameter have only a negligible impact on the amplitude distribution and the mean amplitude-redshift relations predicted by our simulations, and therefore pose no threat to our conclusions.

A more serious concern than the assumed values of our model parameters is a number of simplifying approximations on which the microlensing simulations are based.  

The multiplicative magnification approximation, which despite having been used in similar contexts several times before (e.g. Schneider \cite{Schneider}, Lacey \cite{Lacey}) and having been shown to give reasonable accuracy when it comes to the statistics of magnification distributions (Pei \cite{Pei 1993}), has not been thoroughly tested for the case of light-curve statistics. A particular concern is that it may underpredict the amplitudes, since the high-amplification features associated with caustic crossings are not well-reproduced in this approximation. This can e.g. be seen by comparing the light curves in Schneider (\cite{Schneider}) where the multiplicative magnification approximation is used, with those in Lewis et al. (\cite{Lewis et al.}), which rely on a ray-shooting approach. The naive expectation would be for inclusion of light-curve features associated with caustic crossings to increase the mean amplitudes and flatten the average structure function due to introduction of more short-term variability. The actual effect will however depend on the details of the sampling of the light curves and is hard to quantify without making a direct comparison between the two methods. If a proper treatment of caustic crossings would in fact give a significant contribution to the predicted amplitudes, this would also strengthen the upper limits on $\Omega_\mathrm{compact}$ inferred by Zackrisson \& Bergvall (\cite{Zackrisson & Bergvall}), where a large part of the relevant small source microlensing parameter space has already been ruled out. For this reason, it is doubtful whether relaxing the approximation of multiplicative magnification really can bring the model into better agreement with observations.

Our simulations furthermore assume all compact objects to be randomly and uniformly distributed with constant comoving density throughout the simulated observing beam, an approach commonly referred to as the Press \& Gunn (\cite{Press & Gunn}) approximation. Wyithe \& Turner (\cite{Wyithe & Turner}) showed that although this approximation can give very misleading results if only known stellar populations contribute to the microlensing effect, it is useful for estimating the microlensing optical depth in the case where a significant fraction of the dark matter is in the form of compact objects, as in the scenario explored here. This does not necessarily imply that this approximation is valid when generating light-curve statistics, since correlated lens positions could give rise to correlated magnifications in the simulated time series and therefore affect the distribution of amplitudes. When evaluating such possible effects, one should instead of the isolated dark matter halos assumed in Wyithe \& Turner more realistically use the mass distributions predicted when sending the observing beam through cosmological N-body simulations of large-scale structure. The naive expectation for the inclusion of clustered lenses is however an increased bimodality in the distribution of amplitudes, as some simulated observing beams will contain huge concentrations of compact objects whereas other will be essentially devoid of matter. This is likely to decrease the fraction of light curves with amplitudes above 0.35 magnitudes, since a growing number of quasars should display practically no microlensing variability. The demonstrated discrepancy between microlensing predictions and observations in Fig.~\ref{fracpic} are therefore expected to grow even further as effects of clustering are introduced. 

Testing the reliability of these approximations is not only important for investigations of the present kind, but also for modelling the time-dependent microlensing magnification of other light sources at cosmological distances, like supernovae and the afterglows of gamma-ray bursts. To conduct the required investigation does however represent a huge undertaking and is not within the scope of this article. Instead, detailed tests of the multiplicative magnification approximation and the Press \& Gunn approximation in the context of light-curve statistics are intended to form the basis of a future paper.

Assuming that our simulations really are sufficiently reliable, could it then be that the discrepancy between microlensing predictions and observations is simply due to a missing fraction of low amplitude quasars in the UVX sample? Ultraviolet excess is normally not assumed to correlate with variability (e.g. Hawkins \& V\'eron \cite{Hawkins & Véron}), but provided that the quasar continuum radiation originates from a large accretion disc which becomes bluer towards the center, one could still argue that a lensing event with small impact parameter would statistically both increase the amplitude and decrease $U-B$, thereby introducing a correlation between these two parameters. There are however two reasons why this effect is unlikely to produce a bias of the required strength. Firstly, because the observed achromaticity of the quasar light curves are inconsistent with strong colour gradients across the accretion disc if the variations truly are caused by microlensing in the large source regime, and secondly because huge numbers of quasars would have to be present on the redward side of the $U-B < -0.2$ threshold, which does not appear to be the case (e.g. V\'eron \cite{Véron}).

\section{Summary} 
Using a microlensing model based on the multiplicative magnification approximation and the assumption of uniformly and randomly distributed compact objects, we find that microlensing can reasonably well reproduce the observed average structure function of quasars, but fails to reproduce the high fraction of objects with amplitudes higher than 0.35 magnitudes and the mean amplitudes observed at $z_\mathrm{QSO} < 1$. This indicates that another mechanism must also contribute significantly to the long-term optical variability and severely complicates the task of using light-curve statistics from quasars which are not multiply imaged to isolate properties of any cosmologically significant population of compact objects which may be present. The one piece of evidence pointing uniquely in the direction of an extrinsic variability mechanism is however the lack of cosmological time dilation in the characteristic time scales of the variations. If no other efficient mechanism for extrinsic variability than microlensing can be found, one possible way of setting a lower limit to the contribution from microlensing could therefore be to investigate how large the microlensing population would have to be to eradicate the cosmological time dilation signal believed to be a feature of any intrinsic component of variability.
\begin{acknowledgements}  
We wish to thank M.R.S. Hawkins for useful discussions and for providing us with the average structure function derived from the UVX sample of Hawkins (\cite{Hawkins 2000}). Our referee, L. Wisotzki, is acknowledged for helpful comments on the manuscript. 
\end{acknowledgements}

\end{document}